# Local doping of an oxide semiconductor by voltage-driven splitting of anti-Frenkel defects

Jiali He[1], Ursula Ludacka[1], Kasper A. Hunnestad[1,2], Didrik R. Småbråten[1,3], Konstantin Shapovalov[4], Per Erik Vullum[5], Constantinos Hatzoglou[1], Donald M. Evans[1,3], Erik D. Roede[1], Zewu Yan[6,7], Edith Bourret[7], Sverre M. Selbach[1], David Gao[8,9], Jaakko Akola[8,10] and Dennis Meier[1]

[1]Department of Materials Science and Engineering, NTNU Norwegian University of Science and Technology, Trondheim, Norway.

[2]Department of Electronic Systems, NTNU Norwegian University of Science and Technology, Trondheim, Norway.

[3]Department of Sustainable Energy Technology, SINTEF Industry, Oslo, Norway.

[4]Theoretical Materials Physics, Q-MAT, University of Liège, B-4000 Sart-Tilman, Belgium.

[5]SINTEF Industry, Trondheim, Norway

[6]Department of Physics, ETH Zürich, Zürich, Switzerland.

[7]Materials Sciences Division, Lawrence Berkeley National Laboratory, Berkeley, USA.

[8]Department of Physics, NTNU Norwegian University of Science and Technology, Trondheim, Norway.

[9]Nanolayers Research Computing, London, UK.

[10]Computational Physics Laboratory, Tampere University, Tampere, Finland.

**Layered oxides exhibit high ionic mobility and chemical flexibility, attracting interest as cathode materials for lithium-ion batteries and the pairing of hydrogen production and carbon capture. Recently, layered oxides emerged as highly tunable semiconductors. For example, by introducing anti-Frenkel defects, the electronic hopping conductance in hexagonal manganites was increased locally by orders of magnitude. Here, we demonstrate local acceptor and donor doping in $Er(Mn,Ti)O_3$, facilitated by the splitting of such anti-Frenkel defects under applied d.c. voltage. By combining density functional theory calculations, scanning probe microscopy, atom probe tomography, and scanning transmission electron microscopy, we show that the oxygen defects readily move through the layered crystal structure, leading to nano-sized interstitial-rich (p-type) and vacancy-rich (n-type) regions. The resulting pattern is comparable to dipolar npn-junctions and stable on the timescale of days. Our findings reveal the possibility of temporarily functionalizing oxide semiconductors at the nanoscale, giving additional opportunities for the field of oxide electronics and the development of transient electronics in general.**

Precise control of defects in solids is essential to introduce and tune functional properties. A classical example is p- and n-type semiconductors, which owe their electronic properties to the introduction of acceptor and donor atoms, respectively, representing the backbone of modern information and communications technology[1,2]. Oxide semiconductors are a special sub-category, which offer an outstanding tunability when it comes to defects[3]. In these materials, oxygen defects (i.e., vacancies and interstitials) can readily be used to manipulate the mechanical[4], electric[5], and magnetic properties[6], giving rise to additional degrees of freedom not available in conventional semiconductors. The flexibility of oxides has motivated the design of innovative device architectures for next-generation nanoelectronics and spintronics. Opportunities range from oxide-based memristors for neuromorphic computing[7,8,9] to energy-efficient spin filters and logical gates[10,11].

In recent years, tremendous progress has been made in controlling oxygen defects in oxide semiconductors[12,13,14,15]. An intriguing example are $SrTiO_3$ superlattices with oxygen vacancy gradients, which were achieved by tuning the oxygen pressure during synthesis, demonstrating atomic-scale control[16]. Importantly, oxygen defects are particularly versatile and can be generated and manipulated even after a material has been synthesized. For instance, it has been demonstrated that by annealing $LuFe_2O_4$ in an oxygen-rich atmosphere, oxygen interstitials can be created at will, altering both the material's charge ordering and spin configuration[17]. Furthermore, local manipulation of oxygen defects is possible using a broad range of stimuli, including intense light fields, X-rays, focused electron beams, and electrical voltage[15,18]. Examples are the formation of a two-dimensional electron gas at the surface of $KTaO_3$ under irradiation of intense UV light[19] and the electric-field-driven metal-insulator transition in $VO_2$, which relates to the creation and annihilation of oxygen vacancies[20]. For a broader overview of this rapidly evolving field, the interested reader is referred to recent review articles, e.g., ref.[21] and ref.[22].

In addition to just individual oxygen vacancies ($V_O$) or interstitials ($O_i$), more complex defect arrangements can be applied to control the material's physical properties. Schottky defects in MgO, consisting of paired magnesium and oxygen vacancies, were found to exhibit better stability and a lower migration energy than isolated vacancies[23]. In cubic $Lu_2O_3$, the formation of oxygen vacancy-vacancy pairs introduces additional energy levels within the bandgap, contributing to the coloration and optical absorption properties[24]. Recently, in 0.2% Ti-doped erbium manganite, $Er(Mn,Ti)O_3$, anion interstitial-vacancy pairs (anti-Frenkel defects) were shown to locally enhance the conductance by up to four orders of magnitude, enabling conductivity control at the nanoscale[25].

Motivated by this pronounced correlation between oxygen defects and electronic transport properties in an oxide semiconductor, and its remarkable flexibility when it comes to accommodating oxygen defects in general, we focus on layered hexagonal $Er(Mn,Ti)O_3$ as the model system in this work. In our spatio-temporal experiments, we explore how different combinations of voltage and exposure time can be utilized to inject and control oxygen defects. We demonstrate that nano-sized interstitial-rich (p-type) and vacancy-rich (n-type) regions can be created on demand and develop a microscopic model for the defect formation process.

$Er(Mn,Ti)O_3$ is a ferroelectric p-type semiconductor[26], which has been intensively studied with respect to its multiferroic properties, as well as the emergence of unusual ferroelectric domains and functional domain walls[27,28,29,30,31,32]. Most importantly for this work, the system exhibits an outstanding chemical flexibility that can be leveraged for property engineering via oxygen defects[26,33,34,35,36]. In contrast to perovskite-type oxides, oxygen vacancies and interstitials play an equally significant role in the electronic transport properties of hexagonal manganites[37]. It is established that both n- and p-type behavior can be induced on demand[38], facilitated by the layered and rather open hexagonal crystal structure, where atoms are less densely packed than in perovskite systems.

**Spatio-temporal evolution of electric-field-induced oxygen defects**

We begin with a systematic investigation of changes in the local transport behavior of $Er(Mn,Ti)O_3$ in response to applied d.c. voltages, varying the magnitude and exposure time. To apply the voltage with nanoscale spatial precision and subsequently record conductance maps, a standard conductive atomic force microscopy (cAFM) setup is used as sketched in Fig. 1a, following the same protocol as in ref.[25]: To control the local transport behavior, a conducting AFM tip is brought in contact with the surface and a negative write voltage ($U^{write}$) is applied to the back of the sample ($\approx 0.5$ mm thick, mounted with silver paste on a metal plate) for a certain time ($t^{write}$), whereas read-out is realized using positive voltage ($U^{read}$). Figure 1b displays a representative conductance map gained on a [001]-orientated surface of $Er(Mn,Ti)O_3$ ($U^{read} = +15$ V) after writing a set of conducting features with varying $U^{write}$ (-4.5 V to -16.5 V) and $t^{write}$ (10 s to 30 s). Positions where the writing voltage was applied exhibit bright contrast, indicating an enhancement in conductance by up to one order of magnitude relative to the background. Smaller variations in the background signal ($\Delta I^{domain} \sim 15$ pA) are associated with the ferroelectric domains[31,39].

On a closer inspection of the conductance map in Fig. 1b, we find two characteristic regions. Conducting features written with lower ($U^{write}$, $t^{write}$)-values tend to exhibit a dot-like shape, as seen in the area marked with the white dotted line (Region 1). This result is consistent with literature, and we attribute the enhanced conductance to electric-field-written anti-Frenkel defects[25]. In contrast, for larger ($U^{write}$, $t^{write}$)-values, a more complex pattern appears (Region 2), consisting of a conducting central dot and a conducting outer ring, separated by an intermediary region of higher resistivity.

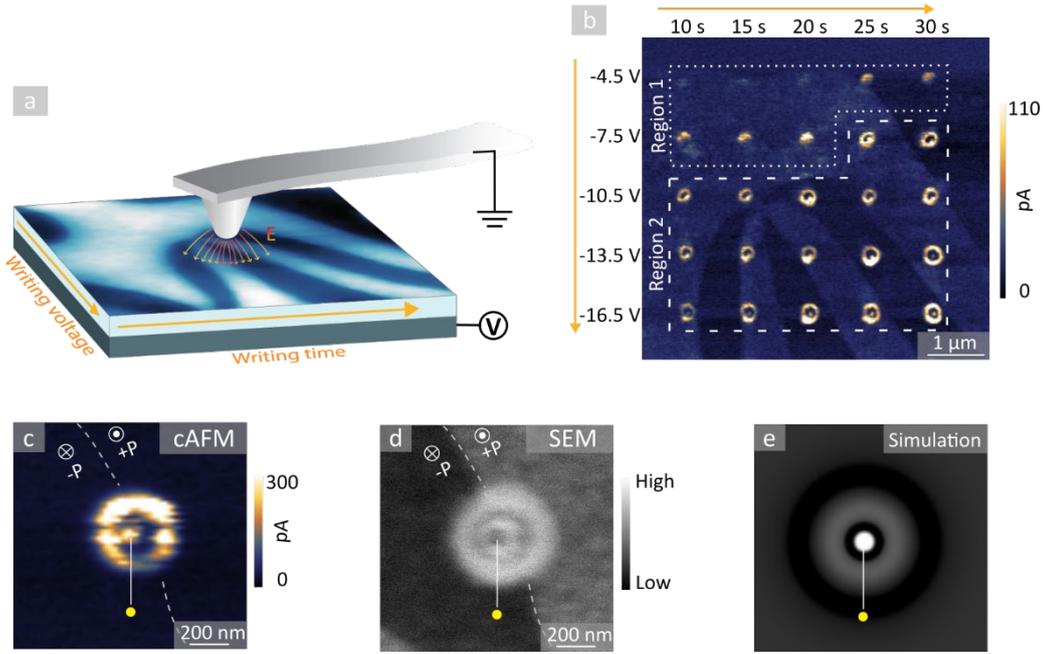

**Fig. 1 | Controlling conductance by electric-field-induced oxygen defects. a**, Illustration of the AFM-based setup used for oxygen defect writing with orange arrows giving the directions in which the writing parameters ($U^{write}$, $t^{write}$) in (**b**) are varied. Negative bias voltages $U^{write}$ are applied to the back electrode. Red/yellow field lines represent the local electric field, $E$, generated by the tip. **b**, cAFM map showing conducting features written in Er(Mn,Ti)$O_3$ on a [001]-oriented surface using different writing parameters ($U^{write}$, $t^{write}$). The image is recorded with a positive voltage $U^{read}$ = +15 V applied to the back electrode, using the same tip as for the defect writing (DCP 20, tip radius 100 nm). **c**, cAFM image of an extended conducting dot-ring structure ($U^{write}$= -25.5V, $t^{write}$= 90 s). White symbols indicate the polarization direction, $P$, of ferroelectric domains in the background. The white dashed line indicates the domain wall position. **d**, SEM image of the same dot-ring structure as in (**c**), captured at 500 V with 0.1 nA electron beam current and 1500 V beam deceleration using the through the lens detector (TLD). **e**, Simulated electric-field-driven distribution of oppositely charged defects. Dark and bright contrasts indicate low and high defect concentrations, respectively (adapted from ref.[25]). The yellow dot in (**c-e**) marks the location of the conducting center of the dot-ring structures.

Figure 1c displays a conductive atomic force microscopy (cAFM) image of an extended dot-ring structure, which was achieved by setting the write time to $t^{write}$= 90 s ($U^{write}$ = -25.5 V). To gain additional insight into the electronic structure, we perform complementary scanning electron microscopy (SEM) measurements on the same feature, allowing for contact-free conductance mapping[40,41]. Figure 1d is an SEM image recorded at the same position as the cAFM map in Fig. 1c, revealing a qualitatively similar dot-ring feature, which corroborates that it is intrinsic to the sample and not related to specifics of the tip-sample contact that co-determines the cAFM conductance map[31].

Compared to the cAFM data in Fig. 1c, however, the contrast in SEM is inverted, i.e., bright conducting regions in cAFM are dark in SEM and vice versa. The same inversion effect is observed for the domain-related contrast in the background, which can be related to the SEM imaging voltage[42,43].

Independent of the inverted contrast, Fig. 1d unveils the existence of an extra outer ring that is not resolved by cAFM. Similar to the insulating intermediary region that separates the two conducting rings, this outer ring is brighter than the domains in the background in the SEM image, indicating that it represents a region with reduced conductance. The SEM data is remarkable as it is in one-to-one agreement with the transport behavior expected based on numerical simulations. For a direct comparison, we show the simulated distribution of $V_O^{\cdot\cdot}$ and $O_i^{''}$ in Fig. 1e (adapted from ref.[25]). The simulation assumes that $V_O^{\cdot\cdot}$ and $O_i^{''}$ are created by the electric field under the AFM tip; they move in opposite directions due to their charge, including the recombination of $V_O^{\cdot\cdot}$ and $O_i^{''}$. Based on the comparison of the SEM data and the simulation, we conclude that the extra outer ring resolved in Fig. 1d relates to a gradient in defect concentration. This gradient results from the migration of $O_i^{''}$ towards the high-field region in the center, leading to a locally reduced defect concentration and, hence, lower conductance. In summary, Fig. 1 demonstrates that by an adequate choice of the writing parameters $U^{\text{write}}$ and $t^{\text{write}}$, the local concentration of oxygen defects can be controlled, leading to a splitting of injected anti-Frenkel pairs $(V_O^{\cdot\cdot}, O_i^{''})$ and a redistribution of $O_i^{''}$, which is the predominant type of oxygen defect in as-grown hexagonal manganites[26].

**Microscopic defect formation mechanism**

Interestingly, the written dot-ring structures in Fig. 1 have a diameter of up to 310 nm, which is more than ten times larger than the diameter of the contact area between the tip and the sample during the defect writing process ($\approx$ 20 nm, Supplementary Note 1). To analyze the defect dynamics and understand the underlying microscopic mechanisms, we perform density functional theory (DFT) calculations for a 3×3×1 model structure of hexagonal YMnO₃ using the Climbing-Image Nudged-Elastic-Band method for migration path modeling (the absence of f-electrons in YMnO₃ simplifies the calculations compared to the structurally and electronically similar system ErMnO₃). Figure 2 displays the formation process of an anti-Frenkel defect and the subsequent migration of oxygen interstitials and vacancies. The defect formation process considered in Fig. 2a and 2b corresponds to a single planar oxygen atom ($O_{p1}$, yellow) jumping to an out-of-plane interstitial position, which leads to a vacancy, $V_{O_{p1}}$, and a small displacement of the nearby planar oxygen ($O_{p2}$, blue). As a consequence, a defect configuration with two off-lattice oxygens $O_i''$ (yellow $O_{i_{p1}}''$ and blue $O_{i_{p2}}''$) arises. Such a configuration with two $O_i''$ at the vacancy edge (marked "u" in the energy profile in Fig. 2b) is weakly metastable with an energy barrier of 0.04 eV protecting it from collapsing upon perturbations. A further displacement of $O_{p1}$ will cause it to reach the initial planar position of $O_{p2}$. Meanwhile, $O_{p2}$ has nudged another planar oxygen off its lattice site and pushed this third planar oxygen ($O_{i_{p3}}''$, green) to a similar interstitial position, creating a new double $O_i''$ configuration farther away ($\approx$ 6 Å) from the vacancy. We consider this the first metastable anti-Frenkel defect configuration. Note that the defect configuration with two interstitial $O_i''$ is different from the common view of anti-Frenkel defects, which refers to only one vacancy and one interstitial $O_i''$. We observe the double $O_i''$ configuration by starting from a single interstitial $O_i''$ and running molecular dynamics at 300 K, where the system turns fast to the energetically more favorable double $O_i''$ configuration. At elevated temperature, i.e., at 1000 K, recombination takes places within a few picoseconds, following the reverse trajectory of the zero-Kelvin-optimized migration path in Fig. 2a.

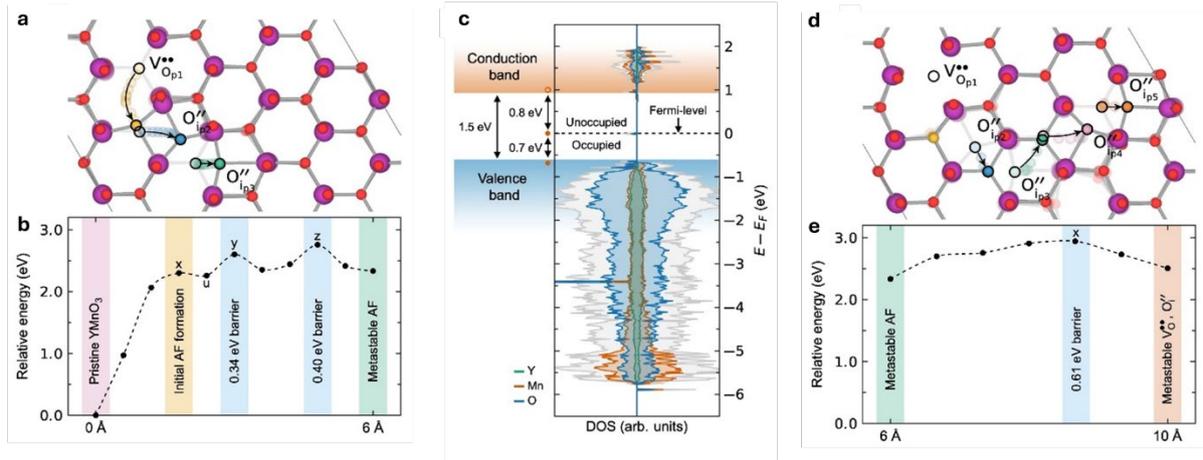

**Fig. 2 | DFT simulations of the oxygen-defect formation. a**, Initial formation pathway for an anti-Frenkel defect and (**b**) its energetics. Local energy maxima along the migration path are labelled x, y, and z. The unstable anti-Frenkel configuration with $O_i''$ at the vacancy edge is labeled "u". **c**, Electronic density of states of $YMnO_3$ with the metastable (6 Å) anti-Frenkel defect configuration. The double $O_i''$ configuration results in an occupied impurity state in the middle of the band gap. **d**, Further migration pathway for the anti-Frenkel defect and (**e**) its energetics. The associated migration barrier, x, is substantially lower than that for the initial defect formation and consistent with that reported for a single $O_i''$ migrating in bulk[26].

Figure 2b shows that the initial vacancy creation step (labeled x, yellow) requires 2.30 eV of energy and that there is only a subtle barrier of 0.04 eV backward. Further migration towards the first metastable anti-Frenkel configuration (green) has two modest barriers of 0.34 eV (labeled y, blue) and 0.40 eV (labeled z, blue). Concerning the stability, the first migration barrier backward, z, is 0.43 eV. The corresponding electronic density of states (DOS, Fig. 2c) shows an impurity state of one electron in the middle of the band gap, which is not present in the defect-free system (Supplementary Fig. S1) and promotes enhanced electronic hopping conductivity[25].

We find that the stability of the defect complex gets further enhanced as oxygen interstitials and vacancies move away from each other. One possible pathway is illustrated in Fig. 2d, showing how $O_{p2}$ and $O_{p3}$ acquire new on-lattice sites by pushing two other oxygen atoms into a double $O_i^{''}$ configuration (pink $O_{i_{p4}}^{''}$ and orange $O_{i_{p5}}^{''}$). In this process, which has a migration barrier of 0.61 eV (labeled x, blue in Fig. 2e), the distance between oxygen interstitials and vacancy grows to 10 Å. A second migration pathway is presented in Supplementary Note 2 and Fig. S1, leading to an energetically less favorable anti-Frenkel defect configuration at 9 Å separation. However, repeating the migration step once more in the same direction (12 Å separation) reduces the energy by 0.44 eV with respect to the previous configuration, indicating improved stability as the distance between double $O_i^{''}$ and $V_{O_{p1}}$ gets larger (not shown).

In summary, our calculations show two important aspects, that is, (i) the system can reduce its energy by separating the oxygen interstitials and vacancies of electric-field-induced anti-Frenkel defects at larger distances, and (ii) there are energy barriers for recombination that lead to different metastable configurations once the driving electric field is switched off. The model indicates a remarkably high mobility of oxygen defects in hexagonal manganites, which is consistent with previous studies[26,44] and the large migration distances we observe experimentally (Fig. 1). Most importantly for this work, the proposed microscopic mechanism, i.e., the electric-field-driven splitting of anti-Frenkel defects, implies that the formation of the dot-ring structures leads to spatially separated $V_O^{\cdot\cdot}$-rich and $O_i^{''}$-rich areas, corresponding to electron- and hole-doped regions, respectively.

**High-resolution imaging of vacancy- and interstitial-rich regions**

To experimentally verify the emergence of such $V_O^{\cdot\cdot}$-rich and $O_i^{''}$-rich regions, we conduct atom probe tomography (APT) measurements. For this purpose, needle-shaped specimens are extracted from two samples using a focused ion beam (FIB) as sketched in Fig. 3a. The first specimen (processed) is taken from an area with an electric-field written conducting dot-ring structure ($U^{write}$ = -52.5 V, $t^{write}$ = 60 s) as sketched in Fig. 3a. The second specimen (pristine) is from an $Er(Mn,Ti)O_3$ reference sample, which is cut from the same crystal and prepared under identical condition as the processed sample (see Methods for details of the sample preparation and Supplementary Fig. S2 for APT mass spectra).

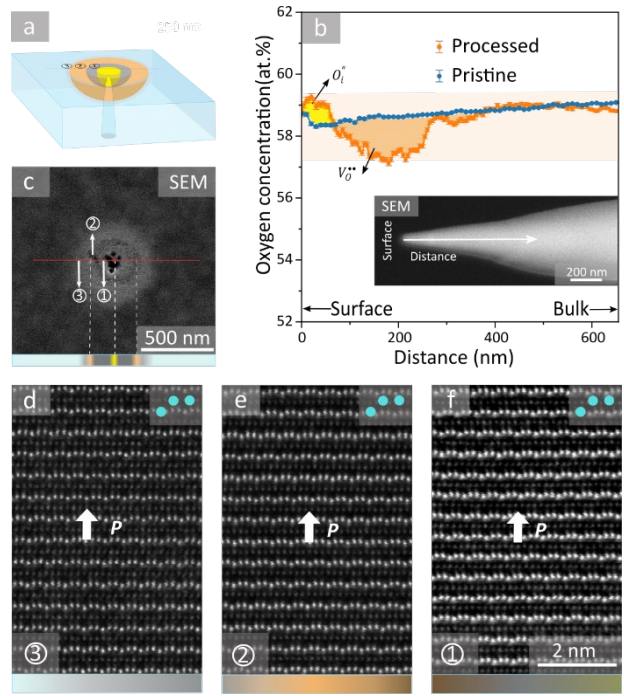

**Fig. 3 | APT and STEM analysis of the dot-ring structure. a,** Illustration of the region from which a needle for APT-based chemical analysis is prepared. Colors and numbers indicate the characteristic regions we resolve by cAFM and SEM (yellow = conducting inner dot; grey = recombination region ①; orange = outer conducting ring ②; light blue = bulk ③. **b,** Oxygen concentration profile measured by APT for an electric-field written dot-ring structure (processed) and an Er(Mn,Ti)O₃ reference sample (pristine). The APT needle extracted from the processed region is displayed in the inset. **c,** SEM image of another dot-ring structure written for subsequent structural analysis ($U^{write}$ = -22.5V and $t^{write}$ = 60 s), captured under an electron beam of 1.25 kV, 0.1 nA, with the TLD detector. The red line marks the location from which a cross-sectional lamella is extracted for HAADF-STEM imaging. **d-e,** HAADF-STEM data for regions ①, ②, and ③.

The needle-like shape enables the high electric fields required for field evaporation of surface atoms in APT as introduced, for example, in ref.[45]. Figure 3b compares the respective oxygen concentration profiles, which are derived by integrating the three-dimensional (3D) APT data perpendicular to the axis of the needle-shaped specimens (see inset to Fig. 3b). We note that a 3.95% offset is added to the Er(Mn,Ti)O₃ reference data to align the curves in the bulk region, accounting for variations in the charge-state ratio (CSR) as discussed in ref.[46] (see Supplementary Fig. S3 and S4 for details). The comparison shows a pronounced variation in oxygen concentration in the electrically modified sample, which is absent in the pristine state. In the electrically modified region, a substantial enhancement in oxygen concentration is measured near the surface (yellow), followed by a region with reduced oxygen concentration (orange). This oxygen concentration profile is consistent with our model and the numerical simulations, and it confirms that the conducting center of the dot-ring structures corresponds to an $O_i^{''}$-rich region, whereas the conducting ring coincides with a $V_O^{\bullet\bullet}$-rich region. However, the width of the intermediary recombination regime obtained based on the APT concentration profile (measured parallel to the [001]-direction) is much smaller than in the cAFM and SEM data (measured perpendicular to the [001]-direction). A possible explanation for this anisotropy is the difference in ionic mobility along and perpendicular to the [001]-axis in hexagonal manganites[26,35,47].

Representative structural data for the characteristic regions ① to ③ as defined for the dot-ring structure in Fig. 3c are displayed in Fig. 3d-f. The SEM data shows qualitatively the same intensity distribution as in Fig. 1d, aside from an additional darker spot in the center where the tip was placed during defect writing, indicating that the high electric-field locally altered the sample (Supplementary Fig. S5). A color bar below the SEM image identifies the different regions of the dot-ring structure, analogous to the illustration in Fig. 3a. The high-angle annular dark-field scanning transmission electron microscopy (HAADF-STEM) data from regions ① to ③ is gained from a cross-sectional lamella extracted at the position marked by the red line in Fig. 3c using FIB (see Supplementary Fig. S5 for details). The HAADF-STEM scans view down the [$\bar{1}00$] direction and exhibit the characteristic down-up–up pattern of Er atoms separated by layers of Mn atoms, reflecting the structural integrity after splitting the anti-Frenkel pairs into $V_O^{\cdot\cdot}$-rich and $O_i^{\prime\prime}$-rich regions. We note that we do not resolve a signature of these defect-rich regions by electron energy loss spectroscopy at room-temperature, which may be attributed to electron-beam induced changes[48], further promoted by a much higher defect mobility compared to the cryogenic temperature at which the APT data in Fig. 3b was recorded (T = 25 K). Importantly for this work, despite the pronounced variation in conductance measurable in cAFM and SEM, the HAADF-STEM data shows qualitatively the same Er displacement within the three regions (down-up-up = +$P$). This finding leads us to the conclusion that the redistribution of oxygen defects is the main driving force for the emergence of the conducting dot-ring structures, whereas potential polarization-dependent contributions can be excluded.

**Transient nature of functionalized regions**

In contrast to the anti-Frenkel defects that arise for lower ($U^{\mathrm{write}}$, $t^{\mathrm{write}}$)-values (Fig. 1b, region 1), the $V_O^{\cdot\cdot}$-rich and $O_i^{\prime\prime}$-rich regions are no longer charge-neutral. As a consequence, different stability criteria are expected to apply in the latter case, which we investigate in Fig. 4. Figure 4a displays a cAFM image gained immediately after writing a conducting dot-ring structure with up to 10 times higher conductance than the background ($U^{\mathrm{write}}$ = 21.0 V, $t^{\mathrm{write}}$ = 90 s). After writing such dot-ring structures, they maintain their initial diameter (here, ≈ 540 nm, red dashed line) and transport behavior with no substantial deviations on the time scale of days. The SEM image in Fig. 1d, for example, was captured six days after the cAFM image in Fig. 1c. On the timescale of months, however, we observe substantial changes. By performing a cAFM scan six months after the initial scan at the same position (Fig. 4a), we find that the well-defined dot-ring structure has vanished. Instead, we now observe a rather blurred spot of only about two to three times higher conductance than the surrounding bulk (Fig. 4b). For a quantitative comparison, line plots extracted along the white dashed lines in the cAFM images in Fig. 4a and 4b are presented in Fig. 4c, highlighting the evolution from a conducting dot-ring structure with three well-defined maxima to a broader region of slightly elevated conductance. The time-dependent behavior of the $V_O''$-rich and $O_i''$-rich regions are thus very different from regions with electric-field-induced anti-Frenkel defects, which were reported to exhibit the robust transport properties even after 24 months[25].

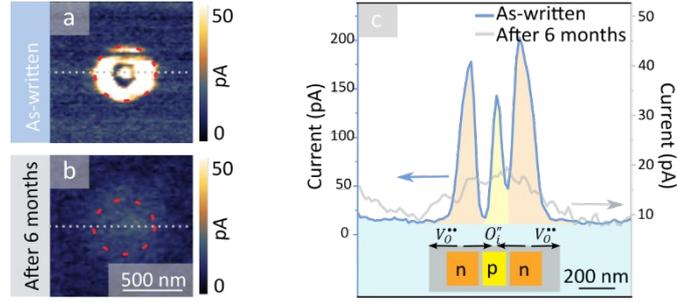

**Fig. 4 | Temporal stability of electric-field written dot-ring structures. a,** cAFM image showing the as-written state of a conducting dot-ring structure. **b,** cAFM image captured at the same position as in (**a**) six months after writing the defect structure. **c,** Current profiles comparing the as-written and the six-months-aged defect structures. The illustration at the bottom depicts the distribution of $O_i''$ and $V_O^{\bullet\bullet}$ associated with the as-written defect structure, transiently generating p- and n-type regimes, comparable to a dipolar npn-junction.

The substantially lower stability compared to the anti-Frenkel defects can be understood based on the positive and negative charges carried by the $V_O''$ and $O_i''$, respectively. Because of the charge of the oxygen defects, the gradients in defect concentration, as observed in Fig. 3b, lead to potential gradients and built-in electric fields that promote the diffusion of charged oxygen defects. Conceptually, the effect is the same as reported in ref.[49,50], where it was observed that oxygen defects collected by a biased tip diffuse and redistribute over time.

**Outlook**

The demonstrated electric-field control of oxygen defects gives new opportunities for the functionalization of oxide semiconductors. Because of the different types of oxygen defects that accumulate at the center ($O_i''$) and in the ring ($V_O''$) of the dot-ring structures studied in this work, different electronic properties arise than in the surrounding material. In particular, $O_i''$ acts as an acceptor and promotes p-type conductivity, whereas $V_O''$ acts as a donor, giving rise to n-type conductivity. Thus, the electric-field-induced defect arrangement can be considered as a bipolar npn-junction as sketched in Fig. 4c. Bipolar npn-junctions are basic electronic components used for amplification and switching[51], enabling the control of currents flowing from the collector to the emitter via a small current injected at the base.

In general, the electric-field-driven splitting of anti-Frenkel defects represents a promising pathway to dope and temporarily functionalize the oxide system, allowing the control of the type of majority carriers with nanoscale spatial precision. The doping relies on oxygen defects and does not enhance the elemental fingerprint, which is of interest for the development of sustainable semiconductor technology. Now that we have shown that interstitial-rich (p-type) and vacancy-rich (n-type) regions are formed under applied d.c. voltage, the next step is to test and utilize the written structures in device-relevant geometries. The results foreshadow conceptually new application opportunities that are of interest for the development of transient oxide nanoelectronics, leveraging the volatile nature of the electric-field-induced nano-regions to locally functionalize a system for a programmed period of time.

**Methods**

**Sample preparation.** Er(Mn,Ti)O₃ single crystals were grown using the pressurized floating-zone method[52]. The samples were then oriented by Laue diffraction and cut into 1 mm-thick pieces with polarization directions perpendicular to the sample surface, yielding specimens with out-of-plane polarization. Following this, the samples were lapped with an Al₂O₃ (9 µm grain size) fluid and polished with silica suspension (Logitech, SF1 polishing suspension) to achieve a smooth surface. After polishing, all samples were annealed in an Entech tube furnace under an N₂ or 5% H₂/N₂ atmosphere at 300 °C for 48 hours, with a heating and cooling rate of 5°C per minute, following the same procedure as in ref. [25].

**Scanning probe and scanning electron microscopy.** cAFM measurements and defect writing were carried out using a Cypher ES environmental AFM (Oxford instruments) equipped with diamond-coated AFM probe tips DCP 20. Bias voltages were applied to the sample back-electrode while the tip was grounded. SEM imaging was performed with a Thermo Fisher Scientific G4UX Dual-beam FIB-SEM operated in secondary electron mode.

**Scanning transmission electron microscopy (STEM).** The lamella for STEM measurements was prepared using Thermo Fisher Scientific G4UX Dual-beam FIB-SEM. A standard extraction procedure was conducted[53] and the milling processes on both sides of the lamella was guided by markers (Supplementary Fig. S5). Low-energy ion beam polishing (2 kV, 0.11 nA) was applied to minimize the beam-induced damage[54]. STEM imaging was performed using a JEOL JEM-ARM200F microscope. The probe size was set to 8C with a beam-limiting aperture of 40 µm, optimized for high-resolution STEM images. A camera length of 4 cm was used for detection, which helps distinguishing heavier atoms in the material through the HAADF detector. This detector is sensitive to Rutherford scattered electrons, and at this camera length, the collection angle starts at 550 mrad, providing enhanced contrast, particularly at the Er sites.

**Atom probe tomography (APT).** APT needles were prepared using a Thermo Fisher Scientific G4UX Dual-beam FIB-SEM. The dot-ring structure was marked by an electron beam deposited carbon (C) marker and followed the extraction procedure described in ref.[55]. APT data were acquired using a Cameca LEAP 5000XS, operating in laser pulsing mode using 30 pJ pulse energy at 250 kHz pulsing frequency. The specimen temperature was set to 25 K, and the detection rate was set to 0.5 % (1 atom detected every 200 pulse on average). Data reconstruction was performed with Cameca APSuite 6, using the dynamic voltage reconstruction method[56], with the initial tip radius determined by SEM and the reconstruction parameters set to achieve correct interatomic plane distances along the 001-pole.

**Density Functional Theory (DFT) calculations.** Detailed information about the DFT calculations, including computational parameters and methods, is provided in Supplementary Note 2.

**Acknowledgements.**

J.H., U.L., and D.M. acknowledge funding from the European Research Council (ERC) under the European Union's Horizon 2020 Research and Innovation Program (Grant Agreement No. 863691). D.M. thanks NTNU for support through the Onsager Fellowship Program and the Outstanding Academic Fellow Program. K.A.H. and D.M. thank the Department of Materials Science and Engineering at NTNU for direct financial support. The Research Council of Norway is acknowledged for the support to the Norwegian Micro- and Nano-Fabrication Facility, NorFab, Project No. 295864, the Norwegian Laboratory for Mineral and Materials Characterisation, MiMaC, project number 269842/F50, and the Norwegian Center for Transmission Electron Microscopy, NORTEM (No. 197405). D.R.S. and S.M.S. acknowledge computational resources for density functional theory simulations provided by Sigma2 - the National Infrastructure for High Performance Computing and Data Storage in Norway through projects NN9264K and NN9259K. S.M.S. and D.R.S. also acknowledge support by the Research Council of Norway through projects 302506 and 275139. K.S. acknowledges support from the European Research Council under the European Union's Horizon 2020 research and innovation program (grant agreement no. 724529), Ministerio de Economia, Industria y Competitividad through grant nos. MAT2016-77100-C2-2-P and SEV-2015-0496, and the Generalitat de Catalunya (grant no. 2017SGR 1506). Z.Y. and E.B. were supported by the US Department of Energy, Office of Science, Basic Energy Sciences, Materials Sciences and Engineering Division under contract no. DE-AC02-05-CH11231 within the Quantum Materials program KC2202. J.A. was supported by the Academy of Finland under project no. 322832 and acknowledges computational resources for DFT simulations at CSC-IT Center for Science, Finland.

## Author contributions

J.H. performed the SPM and SEM experiments, prepared the lamellas for STEM using FIB, assisted with the sample preparation for APT, and analyzed the data. U.L. and P.E.V. carried out the STEM measurements and performed relevant data analysis. K.A.H. prepared the APT needle using FIB, conducted the APT measurements, and analyzed the corresponding data. The work by J.H., U.L., and K.A.H. was done under supervision by D.M.; D.R.S., J.A., D.G., and S.M.S. performed DFT calculations and analyzed the data. K.S. conducted the numerical simulation of the electric field-driven redistribution of defects. D.M.E and E.D.R contributed to discussions on experimental processes. Z.Y. and E.B. provided the materials. D.M. devised and coordinated the project. J.H. and D.M. wrote the manuscript, with support from D.R.S., J.A., and S.M.S. for the DFT-related sections. All authors discussed the results and contributed to the final version of the manuscript.

## Competing interests

The authors declare no competing interests.

## Additional information

Supplementary information

# Local doping of an oxide semiconductor by voltage-driven splitting of anti-Frenkel defects

Jiali He[1], Ursula Ludacka[1], Kasper A. Hunnestad[1,2], Didrik R. Småbråten[1,3], Konstantin Shapovalov[4], Per Erik Vullum[5], Constantinos Hatzoglou[1], Donald M. Evans[1,3], Erik D. Roede[1], Zewu Yan[6,7], Edith Bourret[7], Sverre M. Selbach[1], David Gao[8,9], Jaakko Akola[8,10] and Dennis Meier[1]

[1]Department of Materials Science and Engineering, NTNU Norwegian University of Science and Technology, Trondheim, Norway.

[2]Department of Electronic Systems, NTNU Norwegian University of Science and Technology, Trondheim, Norway.

[3]Department of Sustainable Energy Technology, SINTEF Industry, Oslo, Norway.

[4]Theoretical Materials Physics, Q-MAT, University of Liège, B-4000 Sart-Tilman, Belgium.

[5]SINTEF Industry, Trondheim, Norway

[6]Department of Physics, ETH Zürich, Zürich, Switzerland.

[7]Materials Sciences Division, Lawrence Berkeley National Laboratory, Berkeley, USA.

[8]Department of Physics, NTNU Norwegian University of Science and Technology, Trondheim, Norway.

[9]Nanolayers Research Computing, London, UK.

[10]Computational Physics Laboratory, Tampere University, Tampere, Finland.

## Supplementary Notes

**Supplementary Note 1: Tip-sample contact**

In the Hertz contact model, the probe tip is approximated by a sphere that is in contact with the surface. The radius ($a$) of the contact area between the tip and sample is

$$a^3 = \frac{3FR}{4E^*},$$

where $F$ represents the applied force, $R$ denotes the tip radius, and $E^*$ is the effective elasticity modulus.

$E^*$ can be calculated by [1]

$$\frac{1}{E^*} = \frac{1 - \nu_1^2}{E_1} + \frac{1 - \nu_2^2}{E_2},$$

and $E_1$, $E_2$ are the elastic moduli and $\nu_1$, $\nu_2$ the Poisson's ratios.

The value of $E^*$ is determined to be 119 GPa, based on setting Young's modulus to 220 GPa and Poisson's ratio to 0.277 [2] for both the tip and the sample. For DCP 20 probe tips with a curvature radius of 100 nm, the spring constant was calibrated as 92.68 nN/nm, with a deflection inverse optical lever sensitivity (InvOLS) of 17.73 nm/V and a setpoint of 1V applied. Based on these parameters, the diameter of the contact area is calculated to be ≈20 nm.

**Supplementary Note 2: DFT simulation details and alternative defect migration path**

The electronic structure simulations of the defect migration were performed using the CP2K and VASP program packages [3], applying a combination of the PBE and range-separated PBE0 functionals for the exchange-correlation energy [4,5]. CP2K uses a dual Gaussian–plane wave basis set where the kinetic energy cut-off for the auxiliary plane waves for electron density has been set 600 Ry [6]. A molecularly optimized double-$\zeta$ valence-polarized (DZVP) Gaussian basis [7] was used for all elements with valence configurations Y (4s, 5s, 4p, 4d), Mn (3s, 4s, 3p, 3d) and O (2s, 2p) and together with the Goedecker-Teter-Hutter (GTH) pseudopotentials [8]. The Hartree-Fock exchange component in the hybrid PBE0 functional improves the description of electronic structure for transition metal oxides, and unlike the traditional LDA+U method, PBE0 does not require tuning parameters. Our migration calculations employ a tested approach [9] where PBE0 is used for the 3d-orbitals of Mn (in the spirit of the +U correction) while the rest are treated with PBE. This selection, together with the fact that the computational cost of the Hartree-Fock exchange is reduced by using the auxiliary density-matrix method (ADMM) [10] makes the DFT simulations efficient.

The geometry optimizations were performed by the Broyden-Fletcher-Goldfarb-Shanno (BFGS) algorithm for individual structures and the nudged-elastic-band method (NEB) with the climbing image algorithm [11,12] for modeling the migration paths. The 3×3×1 model structures of hexagonal YMnO$_3$ included 270 atoms with an initial antiferromagnetic ordering of local Mn magnetic moments, and periodic boundary conditions were used in all directions. An extended 3×6×1 model structure (540 atoms) where the system was replicated in one direction was used for investigating the anti-Frenkel defect energetics at longer separations based on structural optimizations of the anti-Frenkel defect configurations. The electronic structure analysis of the representative 3×6×1 anti-Frenkel defect configuration is performed by the projector augmented wave (PAW) method as implemented in VASP [13,14,15] using the PBEsol+U[16] functional with U = 5 eV applied to Mn 3d states. Y (4s, 5s, 4p, 4d), Mn (3s, 4s, 3p, 3d), and O (2s, 2p) were treated as valence electrons, with a plane-wave cut-off energy of 550 eV and a k-point density of $2\times1\times2$.

Inspection of effective (Bader) charges shows that $O_i^{''}$ have reduced negative charges of -1.13e in comparison with -1.28e in the pristine sample. The Mn atom that is bound to both $O_i^{''}$ has an increased cationic charge of +1.90 eV (1.73 in pristine), nominally referred to as $Mn^{4+}$. One of the three Mn that surround $V_O^{''}$ has a reduced charge of +1.37e (nominally $Mn^{2+}$), but otherwise, there are no significant changes in charges elsewhere. The magnetic moments of Mn atoms are consistent with the above observations of charge depletion/accumulation while the initial antiferromagnetic ordering prevails. The charge compensation of the anti-Frenkel defect is comparable to our previous work on the double $O_i^{''}$ configuration [9]. Furthermore, it is consistent with the charge compensation of the constituting $O_i^{''}$ and $V_O^{''}$ in bulk [17,18], with the exception that the anti-Frenkel defect is compensated by oxidizing (reducing) one $Mn^{3+}$ to $Mn^{4+}$ ($Mn^{2+}$) compared to two $Mn^{4+}$ ($Mn^{3+}$) for the isolated bulk defects due to the relative positions of the double $O_i^{''}$ and the $V_O^{''}$ as discussed previously [9]. The changes in effective charges highlight that anti-Frenkel defect formation (migration of O anion) is partially counterbalanced by redistribution of the (negative) charge which reduces the attractive Coulomb interaction between the vacancy and double $O_i^{''}$ configuration.

In comparison to the anti-Frenkel defect migration path presented in the main text, an alternative migration path is displayed in Fig. S 1, and it considers the direction where double $O_i^{''}$ will have the least interaction with the vacancy and its periodic images. As above, the same concerted migration mechanism leads to a detour with two barriers and two minima. The first barrier is 0.88 eV and leads to an anti-Frenkel defect configuration at 0.30 eV (AF2 in Fig. S 1b). After the second barrier of 0.65 eV, the next anti-Frenkel defect state is achieved at 0.86 eV (AF3 in Fig. S 1b). Interestingly, the first anti-Frenkel defect configuration is farther away from the vacancy (10 Å) than the second one (9 Å), while the "dimer" is oriented in a different direction. Both configurations are higher in energy than the starting position (AF at 6 Å), *i.e.,* it costs additional energy to increase the anti-Frenkel defect separation for this process. However, by repeating the detour process once more for an extended $3\times6\times1$ model structure (to accommodate periodic boundary conditions) in the same direction such that the final interstitial-vacancy separation increases to 12 Å reduces the total energy by 0.44 eV with respect to the previous step (9 Å). This gives indications that the situation becomes energetically more feasible at longer distances (>10 Å), as one would expect based on charge screening, and that there are (several) substantial energy barriers for recombination once the driving electric field is switched off.

A comparison of the projected density of states (DOS) of pristine $YMnO_3$ and with the metastable (6 Å) AF configuration is shown in Fig. S 1c.

## Supplementary Figures

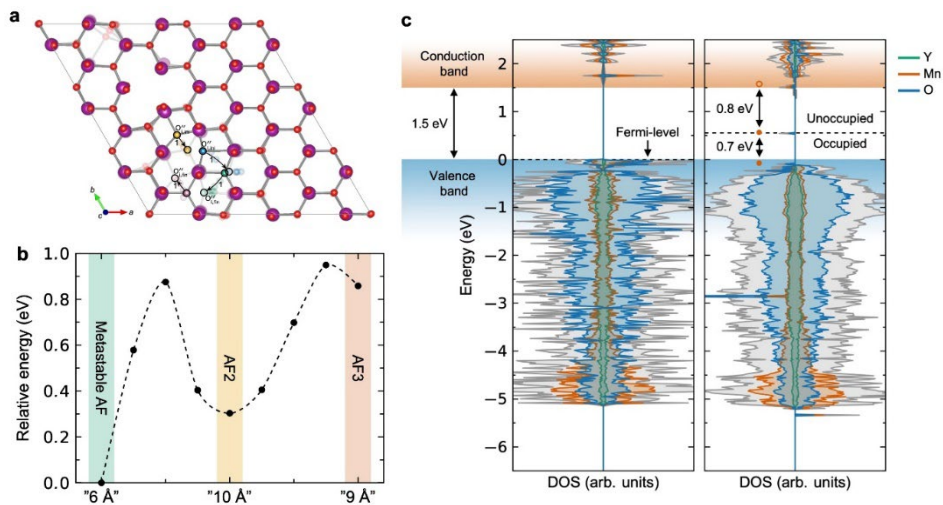

**Fig. S 1 Anti-Frenkel defect migration path. a**, Pathway and (**b**) its energetics. The concerted D-tour migration involves an intermediate double $O_i''$ configuration (blue and green, 0.30 eV) before reaching the final state (green and pink, 0.86 eV). **c**, Comparison of the electronic DOS of pristine YMnO₃ (left) with the metastable (6 Å) anti-Frenkel defect configuration (right). The DOS for the defect system is shifted according to the Fermi level relative to the pristine system.

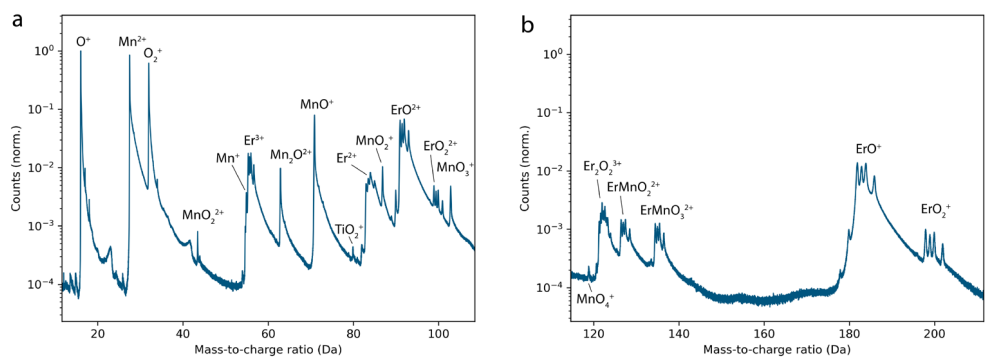

**Fig. S 2 APT mass spectra. a,b**, Mass spectra of the Er(Mn,Ti)O₃ specimen with electric-field written dot-ring structure analyzed in Fig. 3 of the main text. The labeled ionic species indicate the ions used for the reconstruction and the chemical analysis (only the main peaks are labeled, but all isotopes are included). The y-axis is normalized to the largest peak in the spectrum.

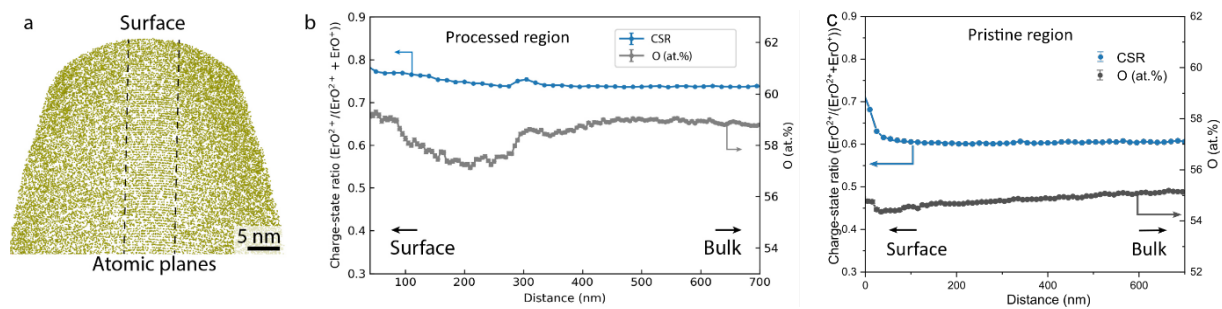

**Fig. S 3 Extended APT analysis. a**, Section of the APT reconstruction of the sample with electric-field written dot-ring structure analyzed in Fig. 3 in the main text. Atomic planes are visible at the 001-pole. Only the Mn ions are used for visualization. **b**, CSR profile for the processed sample, showing a slow decay over the first 400 nm before reaching a stabile value, with a small increase around 300 nm. The minor changes in the CSR (see ref. [19] for a quantitative analysis) do not directly correlate with the observed changes in the oxygen concentration. **c**, CSR profile for the reference sample (pristine). The difference in CSR for the two samples leads to an offset in the measured oxygen concentration, which was accounted for by in Fig. 3b as explained in the main text.

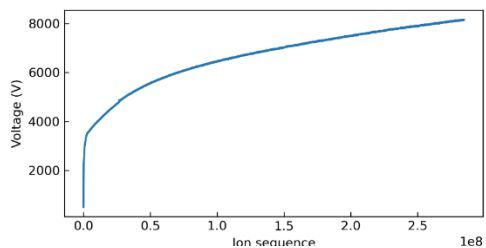

**Fig. S 4 APT voltage profile.** The graph shows the voltage profile used for the dynamic reconstruction method. The voltage increases gradually, indicating a continuously increasing specimen radius as expected during APT analysis. A minor spike in the voltage is observed after 30 M ions around 5 kV, which coincides with the small anomaly in CSR at 300 nm (Fig. S3) and likely corresponds to a microfracture.

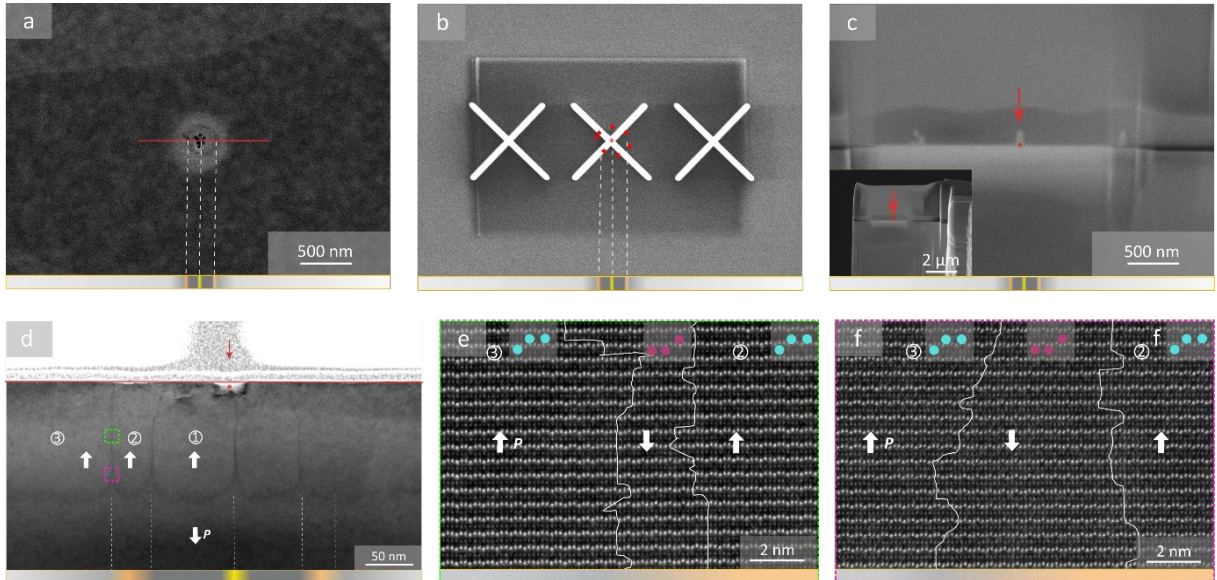

**Fig. S 5 Sample preparation and STEM analysis. a**, Large-scale SEM image of the dot-ring structure presented in Fig. 3c in the main text (1.25 kV, 0.1 nA, TLD). Colors below illustrate the conducting central dot (yellow) and ring (orange) as indicated by the dashed lines. **b**, SEM image of the Pt markers (white crosses; deposited using Pt precursor gas in a low electron beam current, 1.25 kV at 50 pA, in immersion mode) on the dot-ring structure (red dotted line) that were used to extract a cross-sectional lamella from the region of interest. **c**, Cross-sectional SEM image of the lamella; the bright region corresponds to the electron-transparent area [20]. Inset: SEM overview image of the final lamella (4.00 kV, 0.1 nA, ICE detector). Red arrows indicate the position of the marker centered over the dot-ring structure. **d**, Overview STEM image (200 kV, CCD); the red line corresponds to the position marked in **a**. Bright contrast at the top (red dot) indicates local structural changes on the length scale of $\approx 10$ nm, which we attribute to the high electric field at the position where the AFM tip was placed for defect writing (red arrow) [21]. The direction of P is indicated by white arrows, and numbers ①, ②, and ③ label the region for which HAADF-STEM images are presented in the main text. The color bar is a guide to the eye with the same color code as in **a-c**. **e,f**, HAADF-STEM images from the regions marked by green and pink frames in **d**. **e** and **f** show that the darker lines between ①, ②, and ③ in **d** correspond to domains of opposite polarization, pointing in the same direction as in the region at the bottom of **d**. The domain patterns reflect potential electron-beam induced switching [22,23], with a preferred appearance of -P domains in transition regions where the transport behavior changes from insulating to conducting. Most importantly for this work, the data corroborates that the local transport properties do not correlate with the direction of P, excluding a polarization driven mechanism.